\documentclass[twocolumn,showpacs,aps,superscriptaddress]{revtex4-2}
\usepackage{color}
\usepackage{soul}
\definecolor{darkgreen}{rgb}{0,0.6,0}
\definecolor{orange}{rgb}{0.99,0.257,0}
\usepackage{hyperref}

\usepackage{graphicx}
\usepackage{amsmath}
\def\mean#1{\langle#1\rangle}
\newcommand{\lr}[1]{\left\langle #1 \right\rangle}

\newcommand{\om}{\omega}

\newcommand{\epi}{\varepsilon}

\newcommand{\be}{\begin{equation}}
\newcommand{\ee}{\end{equation}}
\newcommand{\ba}{\begin{eqnarray}}
\newcommand{\ea}{\end{eqnarray}}
\newcommand{\bi}[1]{Fig.~\ref{fig:#1}}
\newcommand{\e}[1]{eq.~(\ref{eq:#1})}
\newcommand{\req}[1]{Eq.~(\ref{eq:#1})}

\def\tr{\tau_{\text{ref}}}
\def\eops{\mathrm{e}}
\def\d{\mathrm{d}}
\usepackage{lipsum}

\begin{document}

\title{Fluctuation-response relations and response-response relations for membrane voltage and spike train of stochastic integrate-and-fire neurons}

\author{Kolja Klett}
\affiliation{Bernstein Center for Computational Neuroscience Berlin, Philippstr.~13, Haus 2, 10115 Berlin, Germany}
\affiliation{Physics Department of Humboldt University Berlin, Newtonstr.~15, 12489 Berlin, Germany}

\author{Benjamin Lindner}
\affiliation{Bernstein Center for Computational Neuroscience Berlin, Philippstr.~13, Haus 2, 10115 Berlin, Germany}
\affiliation{Physics Department of Humboldt University Berlin, Newtonstr.~15, 12489 Berlin, Germany}
\date{\today}

\begin{abstract}
Neurons display spontaneous spiking (in the absence of stimulus signals) as well as a characteristic response to time-dependent external stimuli. In a simple but important class of stochastic neuron models, the integrate-and-fire model with Gaussian current noise, both aspects can mathematically be related via fluctuation-response relations (FRRs) as has been shown recently \cite{Lin22}. Here we extend the class of FRRs to include the susceptibilities of the membrane voltage and subthreshold voltage nonlinearity as well as the power spectrum of the membrane voltage. For a simple but often considered IF model, the leaky IF model with white Gaussian noise, we exploit the FRRs and derive explicit expressions for the power spectrum and susceptibility of the subthreshold membrane voltage. We also put forward a relation between the response functions of the spike train and membrane voltage, a response-response relation (RRR) that holds true for a more general setting than considered in most parts of the paper. For the generalized IF model with an adaptation current and colored Gaussian noise we derive an FRR and an RRR. We briefly discuss useful applications of the derived FRRs and RRRs.
\end{abstract}	
 
\maketitle

\section{Introduction}
Nerve cells (neurons) in the brain encode information on sensory signals in sequences of action potentials, called spike trains \cite{KanSch00, RieWar96, DayAbb01}. Often, however, neurons also display a considerable \emph{spontaneous spiking activity} in the absence of sensory stimulation and they also show spiking variability in the presence of stimuli (trial-to-trial variability). Both aspects are indicative of intrinsic and extrinsic noise sources and necessitate stochastic models of neural firing which have by now a long history in computational neuroscience \cite{Hol76, Ric77, Tuc89, GerKis14, GabCox17}. These models take fluctuations into account and can reproduce both the firing statistics of stimulus-free neurons as well as the response and trial-to-trial variability in presence of a stimulus. An important class of phenomenological models that have been applied both to experimental data of single cells as well as in theoretical studies of neural networks is the integrate-and-fire (IF) neuron model \cite{FouBru02, Bur06, Bur06b}.

The calculation of the characteristics of spontaneous firing (e.g. the spike-train power spectrum) and of the response to time-dependent stimulation (e.g. the susceptibility) is a nontrivial problem in the theory of stochastic processes because of the unavoidable nonlinearity in the mathematical description of spike generation. There are a couple of exact results for the power spectrum and susceptibility of integrate-and-fire models with white Gaussian noise \cite{LinLSG02, BruCha01, LinLSG01}, with colored telegraph noise \cite{DroLin17a}, and with white shot noise \cite{RicSwa10, DroLin17}; however, generally, such results are difficult to derive, especially for biophysically more realistic models that include spike-frequency adaptation, short-term plasticity, and/or temporally correlated (colored) noise. 

In statistical physics, it is known that the spontaneous activity and the response to perturbations of stochastic systems are not independent but often connected by fluctuation-dissipation theorems (FDTs), also referred to as fluctuation-response relations (FRRs) \cite{MarPug08}. In recent years, different research groups have attempted to find similar relationships in neural systems by different methods and at very different levels (single neuron, neural population, network of networks). Investigating whole brain regions using electroencephalogram recording and functional magnetic resonance imaging (fMRI), multiple groups found a complex interplay of the spontaneous and evoked activity \cite{He13, HuaZha17}. FDTs were found for symmetrically coupled brain regions in a model of coupled stochastic Stuart-Landau oscillators near criticality fitted to whole-brain fMRI data: the violation of the equilibrium FDT were associated with complex cognitive tasks and used to quantify deviations from equilibrium dynamics in the brain \cite{DecLyn23}. At the population level, the spontaneous fluctuations and response of a linearized noisy version of the Wilson-Cowan model were compared with experimental measurements of spontaneous and evoked activity using magnetoencephalography \cite{SarArv20}. This method has been expanded to account for imbalanced networks in \cite{NanCan23}. For a network of leaky IF neurons in discrete time, a theoretical framework has been proposed based on the probability density of the spike times and been used to derive approximate FDTs. The applicability of this approach to more general time-continuous IF models (which were shown to best describe spiking in real neurons \cite{JolSch08}) appears to be difficult. Further, a recently developed universal description of a multidimensional stochastic oscillator led to yet another simple FRR, although the latter holds only for an abstract complex-valued variable to which the system variables have to be transformed \cite{PerGut23}. 

For stochastically spiking neurons, in particular for the simple yet phenomenologically very successful IF model, it has turned out that spontaneous firing statistics and response characteristics in terms of the spike train and the membrane potential are comparatively simple related. For different types of IF models with Gaussian noise, in \cite{Lin22} one of us derived FRRs that connect the susceptibility of the firing-rate response to the spectral statistics of spike train and subthreshold membrane voltage in the absence of a stimulus signal (i.e. only intrinsic noise is present). By the same approach, FRRs have also been found for IF models with refractory period \cite{PutLin24} and shot noise \cite{StuLin24}. 

In this work, we will use the methods from \cite{Lin22} to derive relations for different IF neurons that focus on the spontaneous and response statistics of the subthreshold membrane voltage, a standard observable in many neuroscientific experiments. In addition to FRRs for the subthreshold membrane voltage of the different models, which we confirm by comparison with stochastic simulations, we will also derive response-response relations (RRRs) that connect the susceptibilities of spike train and membrane voltage. All these relations may turn out to be useful (i) to estimate voltage statistics from spike-train statistics (or the other way around), (ii) to extract intrinsic noise statistics (power spectrum), and (iii) to validate modeling approaches to experimental data involving the IF framework.

Our paper is structured as follows. We start by introducing the IF model as well as typical measures of fluctuations and response characteristics in sec.~\ref{sec:mod_meas}. Then, we derive FRR and RRR for a simple leaky IF model with white Gaussian noise in sec.~\ref{sec:LIF}. In sec.~\ref{sec:adapt}, we extend these relations to the case of a more general IF model with adaptation and colored noise. We conclude by summarizing our main results and discussing possible applications of the derived relations in sec.~\ref{sec:diss}.

\section{Models and measures}\label{sec:mod_meas}
In the IF model the dynamics of the subthreshold membrane potential $v$ are described by a Langevin equation
\be
    \dot{v} = F(v, a) + \eta (t) + \varepsilon s(t) \label{eq:IF_model_general}
\ee
where $F$ captures the deterministic behavior of $v$ and may also include a dynamical adaptation variable $a(t)$ that follows its own dynamics (see below sec.~\ref{sec:adapt}). The noise term $\eta$ models the stochastic influences like channel noise or fluctuating synaptic input. We also include a time-dependent signal $s(t)$ scaled with a small amplitude $\varepsilon$ which will be used below to measure the linear response of the model; we set $\varepsilon = 0$ when the spontaneous activity is modeled. \req{IF_model_general} is not yet complete: to account for the spike generation, the Langevin equation is endowed with a reset rule. Once $v$ reaches a threshold value $v_T$, the neuron is put into a refractory state for an absolute refractory period $\tau_{\text{ref}}$ and afterwards reset to a value $v_R$.

The time instances at which the threshold crossings occur are the spike times $t_i$. The resulting spike train is the sum of Dirac deltas centered at the spike times
\be
    x(t) = \sum_{\{t_i\}} \delta(t-t_i).
\ee
For an illustration of the IF dynamics, see \bi{ts}.
\begin{figure}[ht] 
	\includegraphics[width= 0.484\textwidth,angle=0]{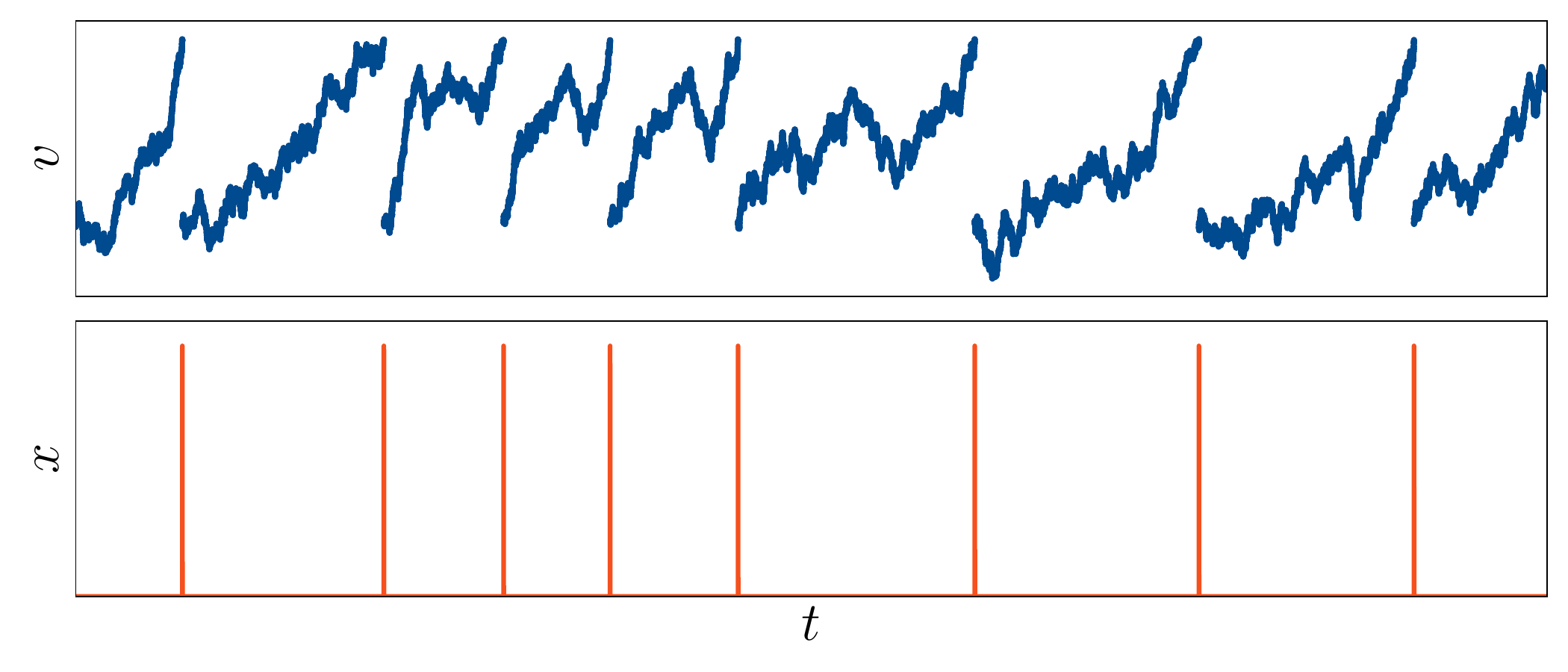}
	\caption{{\bf Time series of the subthreshold voltage and spike train of an IF model.} We use a leaky integrate-and-fire model with $F(v, a) = \mu - v$ and white Gaussian noise $\eta(t) = \sqrt{2D}\xi(t)$ with correlation function $\lr{\xi(t)\xi(t')} = \delta(t-t')$ and noise intensity $D$; here $\mu =0.8$, $D=0.1$, $\tr = 0$ (no refractory period), and $\varepsilon = 0 $ (no input signal).}\label{fig:ts}
\end{figure}    
The fluctuations of a stochastic process $z(t)$ (e.g. the spike train $x(t)$, the membrane potential $v(t)$, or the input signal $s(t)$) can be characterized by its power spectrum $S_{zz}$. For a stationary process, $S_{zz}$ describes how the variance of $z(t)$ is distributed in frequency space. The distribution of the covariance of $z(t)$ and another stochastic process $y(t)$ in frequency space is given by the cross-spectrum $S_{zy}$. Cross-spectrum and power spectrum are related by the Fourier transform to the cross-correlation and autocorrelation functions, respectively, via the Wiener-Khinchin theorem \cite{Gar85}. In numerical simulations, we use Fourier transforms for a finite time window $[0, T]$
\be
    \Tilde{z}(\om) = \int_0^T \d t\, e^{i\om t} z(t).
\ee
The power and cross-spectra are then defined in the limit of an infinite window size by
\be
    S_{zz}(\om)= \lim_{T\to\infty}\frac{\lr{\Tilde{z}(\om)\Tilde{z}^{\ast}(\om)}}{T}, \quad S_{zy}(\om)= \lim_{T\to\infty}\frac{\lr{\Tilde{z}(\om)\Tilde{y}^{\ast}(\om)}}{T}.
\ee
In our simulations we cannot perform this limit but have to choose a sufficiently large window size.

The response of a system to a stimulus $s(t)$ with a small amplitude $\varepsilon$ is described by linear response theory. The time-dependent average of a stochastic process $z(t)$ that attains a stationary value $\lr{z}_0$ in the absence of a stimulus is given by 
\be
    \lr{z(t)} = \lr{z}_0 + \varepsilon \int_{-\infty}^{t} \d t'\, K(t-t') s(t'),\label{eq:LRT}
\ee
where $K(t)$ is the (linear) response function. The latter function quantifies the response to the stimulus but does not depend on $s(t)$. Due to causality, $K(t)$ vanishes for negative arguments, $t < 0$. Hence, the integral in Eq.~\eqref{eq:LRT} is a convolution and has a straightforward representation in frequency space. For non-vanishing frequencies we have
\be
        \lr{\Tilde{z}(\om)} = \varepsilon \chi(\om) \Tilde{s}(\om).
\ee
The Fourier transform of $K$ is called susceptibility and denoted by $\chi$. Since all FRRs derived in this work are in frequency space, we employ $\chi$ as the central response statistics.

\section{Leaky integrate-and-fire model with white noise}\label{sec:LIF}
We consider the simple case of a white-noise-driven leaky integrate-and-fire (LIF) model without adaptation ($F(v,a) = \mu - v$) and refractory state ($\tau_{\text{ref}} = 0$),
\be
    \dot{v} = -v +\mu +\varepsilon s(t)+ \sqrt{2D}\xi(t) - (v_T - v_R) x(t).\label{eq:Langevin_LIF}
\ee
Here, $D$ denotes the intensity of the white Gaussian noise, $\mu$ is the mean input and we have again included a weak stimulus $\varepsilon s(t)$. The last term incorporates in a formal manner the reset rule \cite{LaiCho01, MefBur04, Lin22}. Time is measured in multiples of the membrane time constant $\tau_m$ and voltage is measured in multiples of the threshold-reset distance (see e.g. \cite{VilLin09b}). We note that for $\mu<v_T$ the model operates in the noise-driven or excitable regime, in which there is no spiking without noise or other time-dependent input. In contrast, for $\mu > v_T$, the voltage would repeatedly reach the threshold and be reset even in the absence of any time-dependent driving, which is commonly referred to as the mean-driven regime. The no-spiking limit of the LIF model (achieved by letting the threshold go to infinity, $v_T \to \infty$) is an Ornstein-Uhlenbeck process. 

For this model, even with a non-vanishing refractory period ($\tr > 0$), the stationary firing rate \cite{Ric77}, the spike-train power spectrum \cite{LinLSG02}, and the spike-train susceptibility \cite{BruCha01, LinLSG01} are analytically known and read 
\be
    r_0 = \left( \tr +  \sqrt{2\pi}\int_{z_T}^{z_R}\d z\, \eops^{2z^2} \mathrm{erfc}(\sqrt{2}z)\right)^{-1}\label{eq:r_0_theo}
\ee
(here $z_{T/R}=(\mu-v_{T/R})/\sqrt{D}$), 
\be
S_{xx}(\omega) = r_{0} \frac{\big|\mathcal{D}_{i\omega}(z_T)\big|^2 - \big|\eops^{\frac{z_R^2-z_T^2}{4}}\mathcal{D}_{i\omega}(z_R)\big|^2}{\big|\mathcal{D}_{i\omega}(z_T) - \eops^{i\om\tr}\eops^{\frac{z_R^2-z_T^2}{4}}\mathcal{D}_{i\omega}(z_R)\big|^2},\label{eq:spec-LIF}
\ee
and 
\be
\chi_{x}(\omega) = \frac{i r_{0}\omega/\sqrt{D}}{i \omega - 1}\frac{\mathcal{D}_{i\omega - 1}(z_T) - \eops^{\frac{z_R^2-z_T^2}{4}}\mathcal{D}_{i\omega - 1}(z_R)}{\mathcal{D}_{i\omega}(z_T) - \eops^{i\om\tr}\eops^{\frac{z_R^2-z_T^2}{4}}\mathcal{D}_{i\omega}(z_R)}.\label{eq:susci-LIF}
\ee
In the formulas above, $\mathcal{D}_{a}(z)$ denotes the parabolic cylinder function \cite{AbrSte70}. In what follows, we will set $\tr = 0$ if not explicitly stated otherwise. 

For a vanishing refractory period, spike-train susceptibility and cross- and power spectra of the spontaneous activity are related via the FRR \cite{Lin22}
\be
    \chi_x(\om) =\, \frac{\left(v_T-v_R\right)S_{xx}(\om) + (1+i\om) S_{xv}(\om)}{2D}.\label{eq:xFRR}
\ee

\subsection{Fluctuation-response relation for the membrane voltage} \label{sec:vFRR}
To obtain a FRR for the membrane voltage, we follow the Rice-Furutsu-Novikov scheme suggested in \cite{Lin22}. We start by taking the Fourier transform at non-vanishing frequencies and the complex conjugate of the Langevin \e{Langevin_LIF}. We next multiply the equation with $\Tilde{v}$ and take the average over a stationary ensemble such that we have
\begin{multline}
    (1+i\om) \mean{\Tilde{v}(\om)\Tilde{v}^{\ast}(\om)} =\\ \sqrt{2D}\mean{\Tilde{v}(\om)\Tilde{\xi}^{\ast}(\om)} - (v_T - v_R) \mean{\Tilde{v}(\om)\Tilde{x}^{\ast}(\om)}.\label{eq:Langevin_LIF_FT}
\end{multline}
We apply the Furutsu-Novikov theorem \cite{Fur63}, \cite{Nov65} to express the cross-correlation of $\Tilde{v} $ and $\Tilde{\xi}^{\ast}$ by the response of the mean subthreshold membrane voltage (or, more specifically, we express the covariance by the susceptibility):
\be
    \mean{\Tilde{v}(\om)\Tilde{\xi}^{\ast}(\om)} = \sqrt{2D}\chi_v(\om) \mean{\Tilde{\xi}(\om)\Tilde{\xi}^{\ast}(\om)}.
\ee
Using this relation in \e{Langevin_LIF_FT}, dividing by the observation time $T$, and letting the latter go to infinity, we obtain  
\be
        (1+i\om)S_{vv}(\om) = 2D\chi_v(\om)S_{\xi\xi}(\om) - (v_T - v_R) S_{vx}(\om).
\ee
With the flat power-spectrum of the white noise ($S_{\xi\xi} = 1$), we get
\be
    \chi_v(\om) = \frac{\left(v_T-v_R\right)S_{vx}(\om) + (1+i\om) S_{vv}(\om)}{2D}\label{eq:vFRR}.
\ee

This relation is a FRR: on the left appears $\chi_v$, characterizing the response of the mean subthreshold voltage to an external stimulus ($\varepsilon \neq 0$); on the right we find the spectra $S_{vx}$ and $S_{vv}$, quantifying the spontaneous fluctuations ($\varepsilon = 0$). \req{vFRR} has a similar form as the FRR for the response of the firing rate \req{xFRR}. The difference between \req{vFRR} and \req{xFRR} lays in the occurring response and fluctuation statistics. The new FRR \req{vFRR} is confirmed in \bi{vFRRs} for the noise-driven and mean-driven regime in simulations. A first qualitative observation is that here the voltage response decreases with increasing mean input, an effect that will be discussed in more detail below.

\begin{figure}[ht] 
	\includegraphics[width= 0.484\textwidth,angle=0]{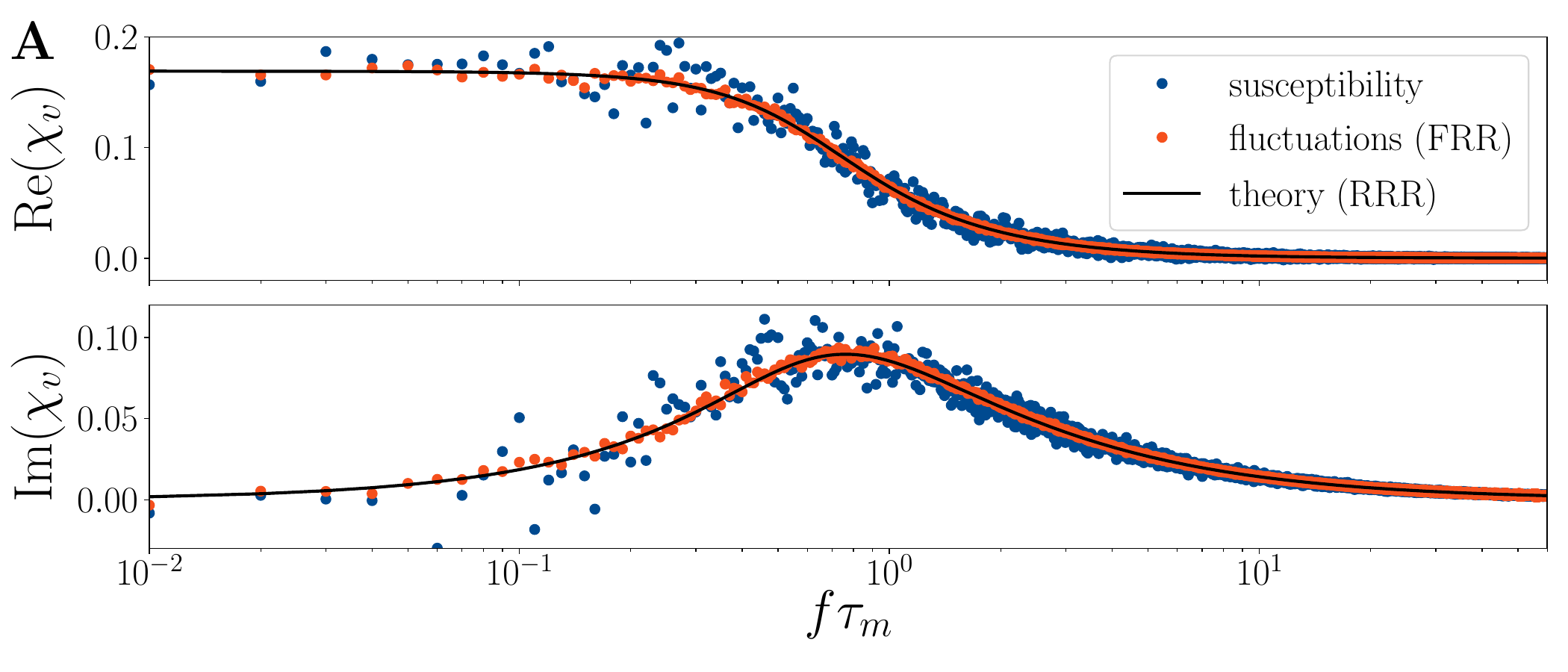}
    \includegraphics[width= 0.484\textwidth,angle=0]{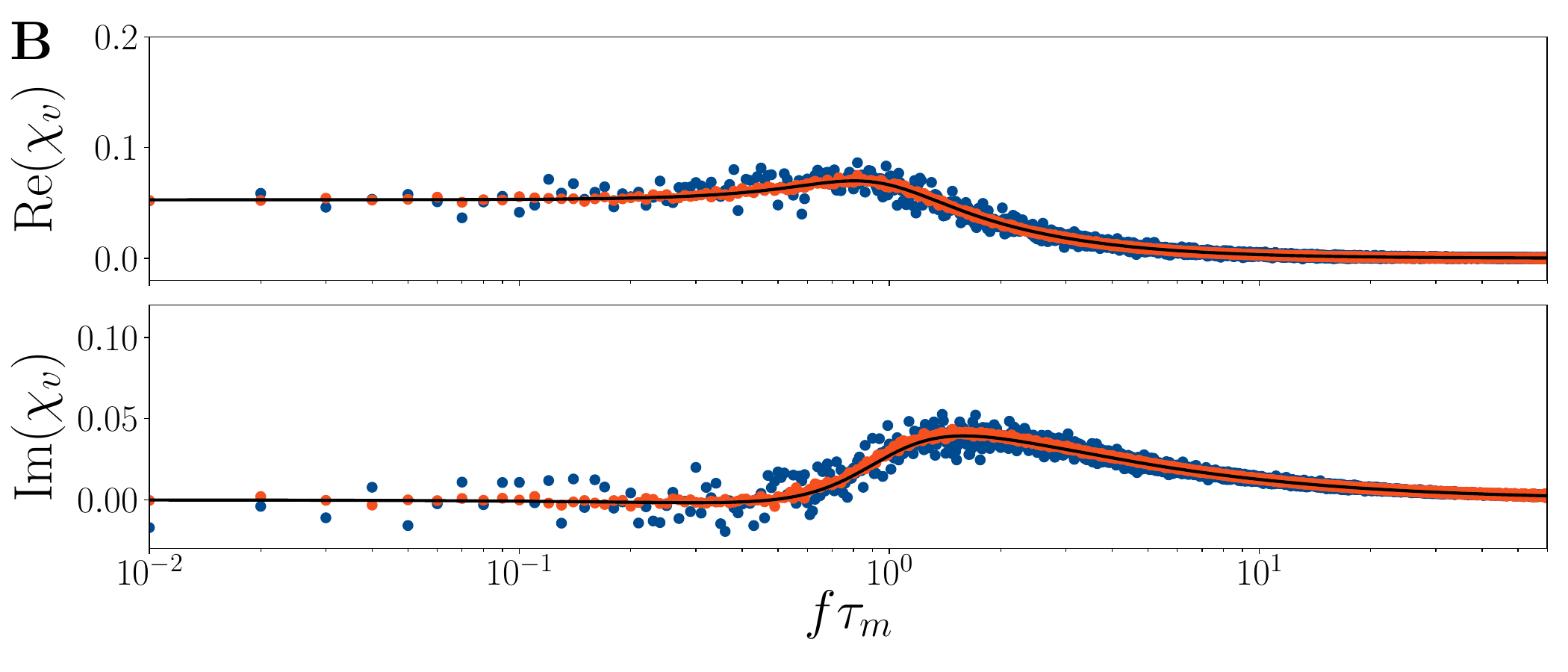}
	\caption{{\bf  Validation of the FRR \req{vFRR} and the RRR \req{RRR}.} Real and imaginary part separately shown in the noise-driven regime (\textbf{A}) and the mean-driven regime (\textbf{B}). For the FRR, we compare simulation data for the response and fluctuation statistics, i.e. left and right of \req{vFRR}, for varying frequency in units of the inverse membrane constant with $D=0.1$, time step of the stochastic simulation $\Delta t = 10^{-4}$, $N = 2\cdot 10^{4}$ realizations and a simulated time window of $T = 100$. The response was determined from stimulation with a broadband noise $\varepsilon s(t)$ with power in the range $[-100, 100]$, scaled such that $s(t)$ has unit variance and $\varepsilon = 1$. For validating the RRR, we compare the simulation data with the analytical result provided by \req{RRR} for $\chi_v$. Remaining parameters: $\mu = 0.8,\ r_0 \approx 0.37 \tau_m^{-1} $ (\textbf{A}) and $\mu = 1.2,\ r_0 \approx 0.73\tau_m^{-1}$ (\textbf{B}).
	}\label{fig:vFRRs}
\end{figure}

\subsection{Response-response relation for voltage and spike train}\label{sec:RRR}
As we demonstrate in this section, there is a simple relationship between the response functions of the subthreshold membrane voltage and that of the firing rate:
\be
\chi_v(\om) = \frac{1 - (v_T - v_R) \chi_x(\om)}{1-i\om}\label{eq:RRR}
\ee
This relation (also confirmed in the numerical tests in \bi{vFRRs}) provides via the knowledge of $\chi_x$ (\req{susci-LIF}) an analytical result for susceptibility of the voltage of an LIF model with white noise. It can be derived in different ways: (i) by the Rice-Furutsu-Novikov method from \cite{Lin22}, (ii) by linear-response theory applied to \req{Langevin_LIF}, and (iii) by linear-response theory applied to the Fokker-Planck equation. So why should we not content ourselves with the simplest way to derive the relation? It turns out that the two alternative derivations hold for more general models. Method (ii) proves that the RRR holds true not only for white-Gaussian noise but for a general (colored or white and Gaussian or non-Gaussian) noise as long as its mean value vanishes. Finally, method (iii) allows to derive a relation for an LIF neuron with white Gaussian noise and an additional refractory period after each spike. Hence, we think that the three methods introduced above have each their own merits and are thus worth to be presented in detail.  

\subsubsection{(i) Derivation using the Rice-Furutsu-Novikov method}
First, we Fourier transform the Langevin \e{Langevin_LIF}, multiply with $\Tilde{\xi}^{\ast}$ and take the average over a stationary ensemble. At non-vanishing frequencies, we get
\begin{multline}
    (1-i\om) \mean{\Tilde{v}(\om)\Tilde{\xi}^{\ast}(\om)} =\\ \sqrt{2D}\mean{\Tilde{\xi}(\om)\Tilde{\xi}^{\ast}(\om)} - (v_T - v_R) \mean{\Tilde{x}(\om)\Tilde{\xi}^{\ast}(\om)}\label{eq:Langevin_LIF_FT_RRR}
\end{multline}
Then, using the Furutsu-Novikov theorem, we replace the averages containing the spike train and the subthreshold membrane potential by the susceptibility of spike train and subthreshold membrane potential, respectively
\ba
    \begin{aligned}
        \mean{\Tilde{v}(\om)\Tilde{\xi}^{\ast}(\om)} &=\sqrt{2D}\chi_v(\om) \mean{\Tilde{\xi}(\om)\Tilde{\xi}^{\ast}(\om)}\\
        \mean{\Tilde{x}(\om)\Tilde{\xi}^{\ast}(\om)} &=\sqrt{2D}\chi_x(\om) \mean{\Tilde{\xi}(\om)\Tilde{\xi}^{\ast}(\om)}.
    \end{aligned}
\ea
The only remaining average in \req{Langevin_LIF_FT_RRR} then is $\mean{\Tilde{\xi}(\om)\Tilde{\xi}^{\ast}(\om)}$ and that is a common factor to all terms in the equation. We can ensure that this common factor does not vanish by dividing by the observation time $T$ and taking $T \to \infty$ resulting in the non-zero power spectrum of the white noise:
\begin{multline}
    (1-i\om) \sqrt{2D}\chi_v(\om) S_{\xi\xi}(\om) =\\ \sqrt{2D}S_{\xi\xi}(\om) - (v_T - v_R) \sqrt{2D}\chi_x(\om) S_{\xi\xi}(\om)
\end{multline}
We cancel $S_{\xi\xi}$ and solve for $\chi_v$ to obtain a RRR between the two response statistics $\chi_x$ and $\chi_v$
\be
    \chi_v(\om) = \frac{1 - (v_T - v_R) \chi_x(\om)}{1-i\om}.
\ee
We can repeat all the steps above for the case of a colored instead of white Gaussian noise, i.e. this derivation also applies for an IF neuron with a correlated Gaussian noise as we have tested in numerical simulations (not shown). The Gaussian property of the noise though is still essential because we invoked the Furutsu-Novikov theorem. 

\subsubsection{(ii) Derivation using linear response in the LIF equation}\label{sec:ii}
Here, we do not rely on the Furutsu-Novikov theorem to introduce the response statistics into the relation. We just apply linear response theory in combination with a certain stimulus.

Consider \req{Langevin_LIF} driven by a complex exponential with amplitude $\varepsilon$ and frequency $\om$
\be
    \dot{v} = -v + \mu + \sqrt{2 D}\xi(t) - (v_T - v_R) x(t) + \epi \eops^{-i\om t}\label{eq:langevin_LIF_eps}.
\ee
Averaging yields an ordinary differential equation for the time-dependent average of $v$
\be
    \frac{\d}{\d t} \mean{v(t)} = -\mean{ v(t)} + \mu - (v_T - v_R) \mean{x(t)} + \epi \eops^{-i\om t}.\label{eq:langevin_LIF_avg}
\ee
The noise drops out as it has zero mean. This is the reason the RRR resulting from this approach holds for a more general class of noise processes. Next, we apply linear response theory to the time-dependent averages of spike train and voltage:
\ba
        \mean{v(t)} &=& \lr{v}_0 + \epi\eops^{-i\om t}\chi_v(\om)\label{eq:LRT_means_v}\\
        \mean{x(t)} &=& r_0 + \epi\eops^{-i\om t}\chi_x(\om)\label{eq:LRT_means_x}
\ea
where $\lr{v}_0$ is the stationary average of the voltage, $r_0$ the stationary average of the spike train, i.e. the stationary firing rate, and $\chi_v$ and $\chi_x$ are the susceptibilities of voltage and spike train, respectively. These relations follow from \req{LRT} for $s(t') = \eops^{-i\om t'}$ and the definition of the susceptibility as the Fourier transform of the response function. 

Inserting \req{LRT_means_v} and \req{LRT_means_x} into \req{langevin_LIF_avg} yields
\begin{align}
    -i\om  \epi \eops^{-i\om t} \chi_v(\om) =& -\epi \eops^{-i\om t}\left[ \chi_v(\om)-(v_T - v_R) \chi_x(\om)+1\right] \notag\\
    &-\lr{v}_0 + \mu - (v_T - v_R) r_0 
\end{align}
Separating by order of $\varepsilon$ gives two relations. In zeroth order, we obtain a relation between the stationary average of the voltage and the stationary firing rate, which was already derived in ref.~\cite{Lin22b}:
\be
    \lr{v}_0 = \mu -(v_T-v_R)r_0.\label{eq:stat_mean_v}
\ee
In first order, we obtain the RRR:
\begin{align}
    \chi_v(\om) = \frac{1 - (v_T - v_R) \chi_x(\om)}{1-i\om}.\label{eq:RRR_ii}
\end{align}
This relation is more general as it is not limited by the requirements of the Furutsu-Novikov theorem (Gaussianity of the noise), i.e. \req{RRR_ii} (thus also \req{RRR}) also holds for non-Gaussian noise such as Poissonian shot noise; we have tested and confirmed the relation for shot noise (not shown).

\subsubsection{(iii) Derivation using the Fokker-Planck equation}
We follow the method by which an analytical result for the response of the firing rate was obtained in terms of confluent hypergeometric functions \cite{BruCha01, FouBru02} or, equivalently and used in the following, parabolic cylinder functions \cite{LinLSG01, LinLSG02, Lin02}. 

Consider the Fokker-Planck equation for the time-dependent probability density $P(v,t)$, corresponding to the LIF model driven by a complex exponential with a small amplitude, \req{langevin_LIF_eps}: 
\begin{align}
    \partial_t P(v,t) =&\ \partial_v \left( v - \mu -\epi\eops^{-i\om t} + D \partial_v \right)P(v,t)\notag\\
    &+ r(t)\delta(v-v_R).\label{eq:FPE}
\end{align}
The source term at the reset voltage in the second line and an absorbing boundary condition at the threshold voltage reflect the reset rule. Furthermore, the probability density vanishes for $v \to -\infty$ and we require continuity at the reset voltage. All these conditions read
\be
    \begin{aligned}
        &\lim_{v\to-\infty} P(v,t) = 0\\
        &\lim_{\epsilon \to 0} P(v_R - \epsilon,t) - P(v_R + \epsilon,t)= 0\\
        &P(v_T,t) = 0.
    \end{aligned}
    \label{eq:bc}
\ee
Due to the time-dependence of the stimulus, \req{FPE} has no stationary solution. However, in the long-time limit the probability density approaches a cyclo-stationary solution with a time dependence that exclusively is due to the stimulus \cite{Jun93, FouBru02}, suggesting the following linear-response ansatz
\be
    P(v,t) = P_0(v) + \epi\eops^{-i\om t}\eops_-(v)\eops_+(v_R) q(v)\label{eq:lin_resp_ansatz}.
\ee
Here, $P_0$ is the stationary density for $\varepsilon = 0$, $q$ describes the cyclo-stationary modulation of the density and $\eops_{\pm}$ are abbreviations for Gaussian functions that simplify the calculations
\be
    \eops_{\mp}(v) = \mathrm{exp}\left [\mp\frac{(v-\mu)^2}{4D}\right ].
\ee
We obtain a formula for the response of the mean subthreshold voltage by combining \req{LRT_means_v} and $\mean{v(t)}$ expressed through an integral over $P$
\begin{align}
    &\lr{v(t)} = \int_{-\infty}^{v_T} \d v\, v\,P(v,t)\notag\\
    &= \int_{-\infty}^{v_T} \d v\, v\, P_0(v) + \epi\eops^{-i\om t}\eops_+(v_R)\int_{-\infty}^{v_T} \d v\, \eops_-(v)q(v)\label{eq:mean_pdf_ansatz},    
\end{align}

where we used \req{lin_resp_ansatz}. Comparing \req{LRT_means_v} and \req{mean_pdf_ansatz}, we obtain 
\be
    \chi_v(\om) = \eops_+(v_R)\int_{-\infty}^{v_T} \d v\, v\, \eops_-(v) q(v)\label{eq:chi_v_int}.
\ee
The main part of the derivation of the RRR using the Fokker-Planck equation is to calculate this integral. To that end, we still need to specify $q$. The function $q$ can be obtained from an ordinary differential equation that results from inserting \req{lin_resp_ansatz} into \req{FPE} and taking only the linear terms in $\varepsilon$ (see \cite{Lin02} and equivalently \cite{FouBru02}). With the solution for $q$ (\cite{Lin02} p. 126) \req{chi_v_int} reads
\begin{align} 
     \chi_v(\om)=&\ \chi_x(\om) \eops_+(v_R)I(v_R) \notag\\
     &- \int_{-\infty}^{v_T}\d v_S\, P_0'(v_S) e_+(v_S)I(v_S), \label{eq:chi_v_I1}
\end{align}
where the function $I(v_\ell)$ ($\ell = R, S$) is given by 
\begin{align}
    I(v_\ell)=&\ \frac{1}{Dq_1(v_T)} \Bigg[q_1(v_\ell)q_2(v_T)\int_{-\infty}^{v_T} \d v\, v\, \eops_-(v) q_1(v) \notag\\
    &-q_1(v_T)q_2(v_\ell) \int_{-\infty}^{v_\ell} \d v\, v\, \eops_-(v) q_1(v) \notag\\
    &- q_1(v_\ell)q_1(v_T)\int_{v_\ell}^{v_T} \d v\, v\, \eops_-(v) q_2(v)\Bigg],\label{eq:I_def}
\end{align}
and $q_1$, $q_2$ are parabolic cylinder functions \cite{AbrSte70}:
\be
    \begin{aligned}
        q_1(v) &= U\left(-\frac{1}{2}-i\om, \frac{\mu-v}{\sqrt{D}}\right),\\
        q_2(v) &= \sqrt{\frac{\pi}{2}}\, V\left(-\frac{1}{2}-i\om, \frac{\mu-v}{\sqrt{D}}\right),
    \end{aligned}
    \label{eq:U_V}
\ee
where the prefactor in the second line is a consequence of setting the Wronskian of the two solutions to unity. Note that we use a different notation for the parabolic cylinder functions than used in \req{spec-LIF} and \req{susci-LIF}. Using the properties of the parabolic cylinder functions, we can evaluate the integrals involving $q_1$, $q_2$ in $I$ (see sec.~\ref{sec:Appen} for details) and we obtain
\begin{align}
    I(v_\ell) =& \, \frac{1}{i\om(1-i\om)}\bigg[\eops_-(v_T)\frac{q_1(v_\ell)}{q_1(v_T)}(\mu-i\om v_T) \notag\\
    &- \eops_-(v_\ell) (\mu-i\om v_\ell)\bigg].\label{eq:I_calc}
\end{align}
Inserting \req{I_calc} back into \req{chi_v_I1} yields
\begin{align}
    &\chi_v(\om) =  \frac{1}{i\om(1-i\om)}\Bigg\{\chi_x(\om)\bigg[i\om v_R - \mu\notag\\ 
    &+\eops_+(v_R)\eops_-(v_T)\frac{q_1(v_R)}{q_1(v_T)}(\mu-i\om v_T) \bigg] \notag\\
    &-\Bigg[\frac{\mu - i\om v_T}{q_1(v_T)}\eops_-(v_T)\int_{-\infty}^{v_T}\d v_S\, P_0'(v_S) \eops_+(v_S)q_1(v_S)\notag\\
    &- \int_{-\infty}^{v_T}\d v_S\, P_0'(v_S)(\mu-i\om v_S)\Bigg]\Bigg\}\label{eq:chi_v_I_1}.
\end{align}
The evaluation of the two remaining integrals is straightforward. The first integral appears in a general expression for the rate susceptibility (Lindner, unpublished result) which can be obtained from the calculational approach presented above and reads
\be
   \chi_x(\om) = \frac{ \eops_-(v_T)\int_{-\infty}^{v_T}\d v_S\ P_0'(v_S) \eops_+(v_S)q_1(v_S)}{\eops_-(v_T)\eops_+(v_R)q_1(v_R)-q_1(v_T) }\label{eq:GEN_SUS}. 
\ee
Hence, we can express by the rate susceptibility $\chi_x$ as follows
\begin{multline}
    \eops_-(v_T)\int_{-\infty}^{v_T}\d v_S\ P_0'(v_S) \eops_+(v_S)q_1(v_S) = \\\chi_x(\om)\left[\eops_-(v_T)\eops_+(v_R)q_1(v_R)-q_1(v_T) \right]\label{eq:SUS_INT}.
\end{multline}
The second integral in \req{chi_v_I_1} is solved using integration by parts, the normalization of $P_0$, and the boundary conditions of \req{FPE} that also hold for $P_0$:
\begin{align}
    \int_{-\infty}^{v_T}\d v_S\, P_0'(v_S)(\mu-i\om v_S) =&\ P_0(v_S) (\mu-i\om v_S)\Big|_{-\infty}^{v_T} \notag\\
    &+ i\om \int_{-\infty}^{v_T}\d v_S\, P_0(v_S)\notag\\
    =&\ i\om\label{eq:P_0_INT}
\end{align}
Inserting \req{SUS_INT} and \req{P_0_INT} into \req{chi_v_I_1}, some terms cancel and we arrive at the RRR
\begin{align}
    \chi_v(\om) = &\, \frac{\chi_x(\om)}{i\om(1-i\om)} \bigg[\eops_+(v_R)\eops_-(v_T)\frac{q_1(v_R)}{q_1(v_T)}(\mu-i\om v_T)\notag\\
    &-\mu + i\om v_R + \mu - i\om v_T \notag\\
    &- e_+(v_R)e_-(v_T)\frac{q_1(v_R)}{q_1(v_T)}(\mu - i\om v_T)\bigg] + \frac{1}{1-i\om} \notag\\
    = &\ \frac{1 - (v_T - v_R) \chi_x
(\om)}{ 1- i\om}.\label{eq:RRR_iii}
\end{align}
While this approach is the most involved requiring solving the Fokker-Planck equation first, it has an advantage compared to the other two. Namely, only a few modifications are necessary to extend the RRR to the case of a non-vanishing refractory period ($\tr \neq 0$) as we show next.

The refractory period enters the Fokker-Planck equation as a delay in the source term at the reset voltage
\begin{align}
    \partial_t P(v,t) =&\ P(v,t) + (v - \mu - \epi\eops^{-i\om t})\partial_v P(v,t)\notag\\
    &+ D \partial_v^2 P(v,t) + r(t-\tr)\delta(v-v_R).\label{eq:FPE_ref}
\end{align}
The linear-response ansatz for $P$ and thus \req{chi_v_int} are not affected by this modification of the Fokker-Planck equation. Only $q$ is endowed with an additional exponential factor such that \req{chi_v_I1} reads
\begin{align} 
     \chi_v(\om)=&\ \chi_x(\om)\eops^{i\om\tr} \eops_+(v_R)I(v_R) \notag\\
     &- \int_{-\infty}^{v_T}\d v_S\, P_0'(v_S) e_+(v_S)I(v_S).
\end{align}
Crucially, $I$ remains unchanged and \req{I_calc} still holds. Hence, \req{chi_v_I_1} is also only endowed with an exponential factor:
\begin{align}
    &\chi_v(\om) =  \frac{1}{i\om(1-i\om)}\Bigg\{\chi_x(\om)\eops^{i\om\tr}\bigg[i\om v_R - \mu\notag\\ 
    &+\eops_+(v_R)\eops_-(v_T)\frac{q_1(v_R)}{q_1(v_T)}(\mu-i\om v_T) \bigg] \notag\\
    &-\Bigg[\frac{\mu - i\om v_T}{q_1(v_T)}\eops_-(v_T)\int_{-\infty}^{v_T}\d v_S\, P_0'(v_S) \eops_+(v_S)q_1(v_S)\notag\\
    &- \int_{-\infty}^{v_T}\d v_S\, P_0'(v_S)(\mu-i\om v_S)\Bigg]\Bigg\}\label{eq:chi_v_I_1_REF}.
\end{align}
The two integrals in \req{chi_v_I_1_REF} are the same as in the non-refractory case. Here, however, they yield different results. The equivalent version of \req{GEN_SUS} in the refractory case reads 
\be
   \chi_x(\om) = \frac{ \eops_-(v_T)\int_{-\infty}^{v_T}\d v_S\ P_0'(v_S) \eops_+(v_S)q_1(v_S)}{\eops^{i\om\tr}\eops_-(v_T)\eops_+(v_R)q_1(v_R)-q_1(v_T) }\label{eq:GEN_SUS_REF}
\ee
and permits to simplify the first integral as follows
\begin{multline}
    \eops_-(v_T)\int_{-\infty}^{v_T}\d v_S\, P_0'(v_S) \eops_+(v_S)q_1(v_S) =\\ \chi_x(\om)\left[\eops^{i\om\tr}\eops_-(v_T)\eops_+(v_R)q_1(v_R)-q_1(v_T) \right]\label{eq:SUS_INT_REF}.
\end{multline}
The calculation of the second integral yields an additional term that combines the spontaneous firing rate and the refractory period:
\begin{align}
    \int_{-\infty}^{v_T}\d v_S\, P_0'(v_S)(\mu-i\om v_S) =&\ P_0(v_S) (\mu-i\om v_S)\Big|_{-\infty}^{v_T} \notag\\
    &+ i\om \int_{-\infty}^{v_T}\d v_S\, P_0(v_S)\notag\\
    =&\ i\om \left[1 - r_0 \tr\right]\label{eq:P_0_INT_REF}.
\end{align}
The additional term stems from the normalization of $P_0$. For $\tr \neq 0$, we have to account for the probability to be in the refractory state 
\begin{align}
    1 = \int_{-\infty}^{v_T}\d v_S\, P_0(v_S) + r_0\int_{t-\tr}^{t}\d t'.
\end{align}
Inserting \req{SUS_INT_REF} and \req{P_0_INT_REF} into \req{chi_v_I_1_REF}, we obtain the RRR for the LIF model with white noise and a non-vanishing refractory period
\begin{align}
    \chi_v(\om)= \frac{1 - r_0\tr - \left[v_T - v_R\eops^{i\om\tr} + \mu \Tilde{B}_{\tr}(\om)\right] \chi_x(\om)}{ 1- i\om}\label{eq:RRR_ref},
\end{align}
where
\begin{align}
    \Tilde{B}_{\tr}(\om) = \frac{\eops^{i\om\tr}-1}{i\om}
\end{align}
is the Fourier transform of a box-car function and we have to use \req{susci-LIF} that includes a dependence on the refractory period $\tr$. Compared to \req{RRR_iii}, the RRR for $\tr \neq 0$ contains some additional frequency-dependent and constant terms. In particular, the mean input appears as an explicit parameter in the relation while for $\tr = 0$ it only appears implicitly through $\chi_x$. Note, $\chi_v$ describes only the response of the subthreshold voltage, as the Fokker-Planck equation contains no information about the voltage during the refractory period. We have tested an LIF model with white Gaussian noise and a non-vanishing refractory period and confirmed the more general RRR \req{RRR_ref}, see \bi{RRR_ref}. The difference to the case of a vanishing refractory period (gray dashed lines) is particularly pronounced at low frequencies. 

\begin{figure}[ht] 
	\includegraphics[width= 0.484\textwidth,angle=0]{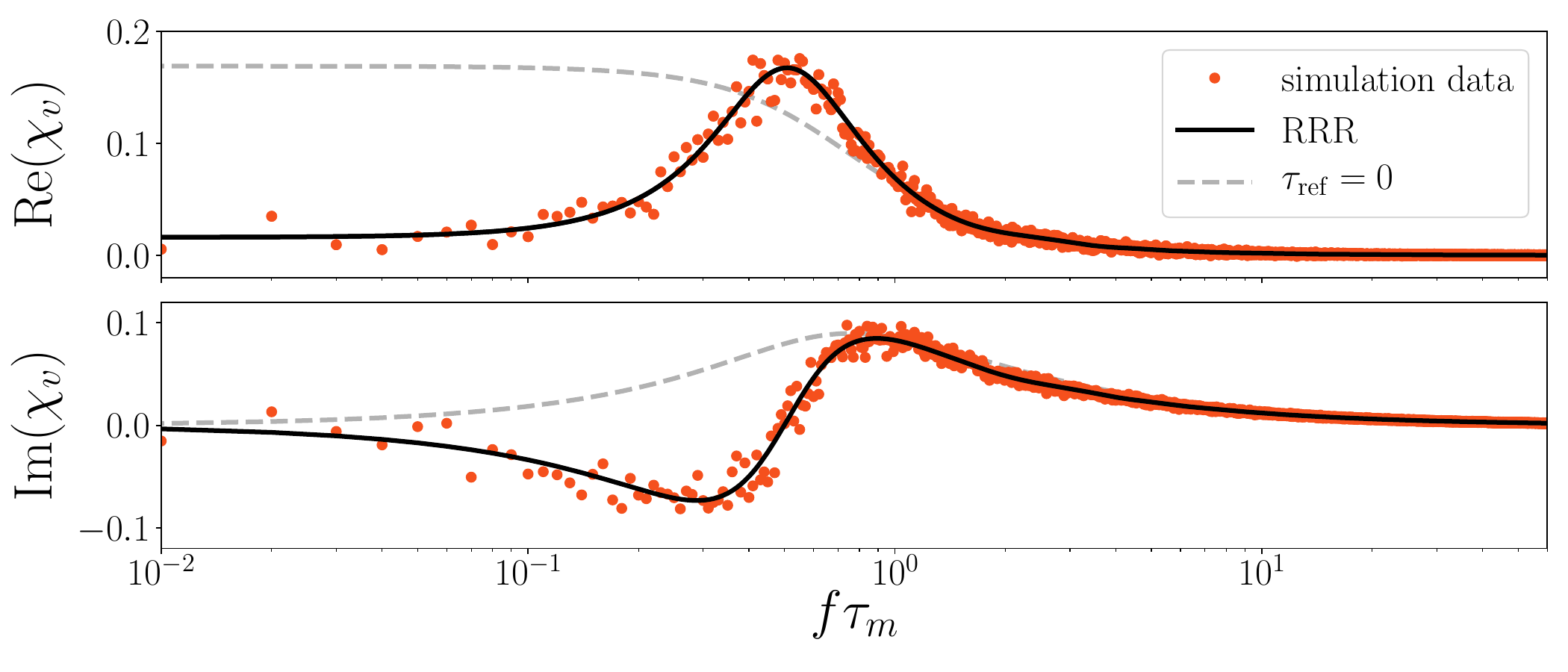}
	\caption{{\bf Validation of the RRR for a non-vanishing refractory period, \req{RRR_ref}.}
    Comparing $\chi_v$ from simulations (dots) and $\chi_v$ from \req{RRR_ref} using \req{susci-LIF} for $\chi_x$ (solid line) for mean input $\mu = 0.8$, noise intensity $D= 0.1$, and refractory period $\tr = 0.5 \tau_m < 1/r_0 \approx 3.2 \tau_m$. The response was determined from stimulation with a broadband noise $\varepsilon s(t)$ with power in the range $[-100, 100]$, scaled such that $s(t)$ has unit variance and $\varepsilon = 1$. The dashed gray line corresponds to $\chi_v$ in the non-refractory case from \bi{vFRRs}(\textbf{A}).}
    \label{fig:RRR_ref}
\end{figure}  

\subsubsection{Comparison of the susceptibilities}
The RRR constitutes a connection between the firing-rate susceptibility and the susceptibility of the subthreshold membrane voltage. For non-weak noise this typically implies that if one susceptibility is large, the other one is small, an effect which is illustrated in \bi{RRR} by a variation of the mean input $\mu$. Without input ($\mu = 0$), the LIF model spikes only rarely and the membrane voltage is close to an Ornstein-Uhlenbeck process (equivalent to \req{Langevin_LIF} without reset term). Consequently, the low-pass-filter susceptibility of this linear process (black line in \bi{RRR} A)
\be
    \chi_{\mathrm{OUP}}(\om) = \frac{1}{1-i\om}\label{eq:susci_OUP}
\ee
is a good approximation for $\chi_v(\om)$ (red line); the spike-train susceptibility (blue line) is low for $\mu = 0$, related to the low firing rate. For an intermediate value of the input ($\mu = 0.5$), the firing rate $r_0$ and the spike-train susceptibility are enlarged and the latter becomes comparable in magnitude to $\chi_v(\om)$ (\bi{RRR} B). Finally, for a strong mean input ($\mu = 1.2$), $\chi_x(\om)$ clearly exceeds $\chi_v(\om)$ (\bi{RRR} C).

\begin{figure}[ht] 
	\includegraphics[width= 0.484\textwidth,angle=0]{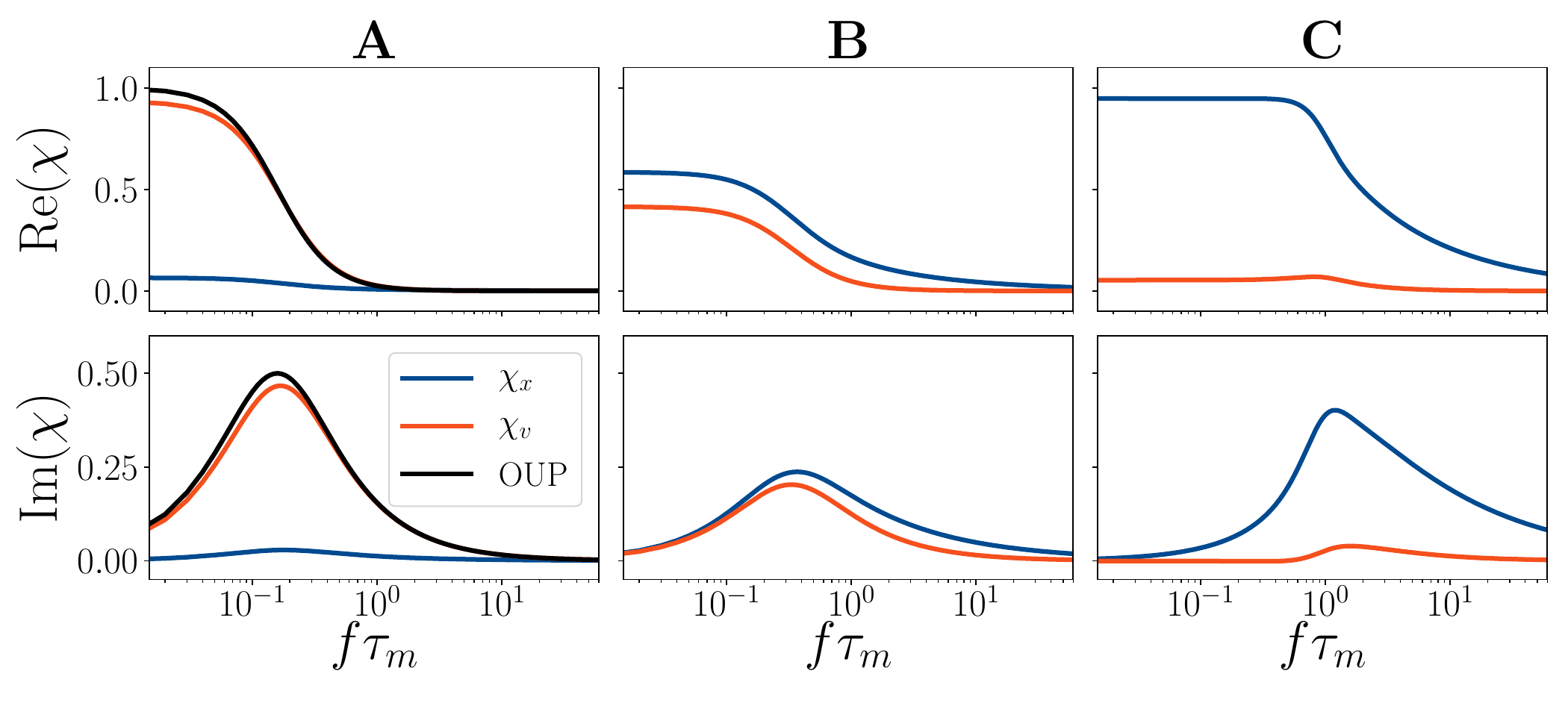}
	\caption{{\bf Comparison of $\chi_x$ and $\chi_v$.}
    Analytical results for $\chi_x$ (\req{susci-LIF}) and $\chi_v$ (\req{RRR} using the known result for $\chi_x$). The mean input $\mu$ and, consequently, the firing rate $r_0(\mu)$ vary across the panels: (\textbf{A}) no input, i.e. $\mu = 0,\ r_0 \approx 0.01 \tau_m^{-1} $; (\textbf{B}) noise-induced firing regime with $\mu = 0.5,\ r_0 \approx 0.15 \tau_m^{-1}$; and (\textbf{C}) mean-driven regime with $\mu = 1.2,\ r_0 \approx 0.73 \tau_m^{-1}$. In all panels, the noise intensity is $D = 0.1$.}
    \label{fig:RRR}
\end{figure}  

\subsection{Power spectrum of the subthreshold membrane voltage}
We can combine the relations derived in this work (\req{vFRR} and \req{RRR}) with \req{xFRR} to express the power spectrum of the subthreshold membrane voltage in terms of statistics that are analytically known, leading to an analytical results for $S_{vv}$:
\begin{align}
    S_{vv}(\om) = \frac{(v_T-v_R)^2 S_{xx}(\om) + 2D\left[ 1 - 2(v_T -v_R)\text{Re}\chi_x(\om)\right]}{1+\om^2}\label{eq:S_vv_theo}.
\end{align}
It is confirmed by simulations in the noise-driven and mean-driven regime, see \bi{S_vv}. For non-weak noise, increasing the mean input reduces the power at low frequencies significantly while a small maximum appears at intermediate frequencies in the mean-driven regime. 

\begin{figure}[ht] 
	\includegraphics[width= 0.484\textwidth,angle=0]{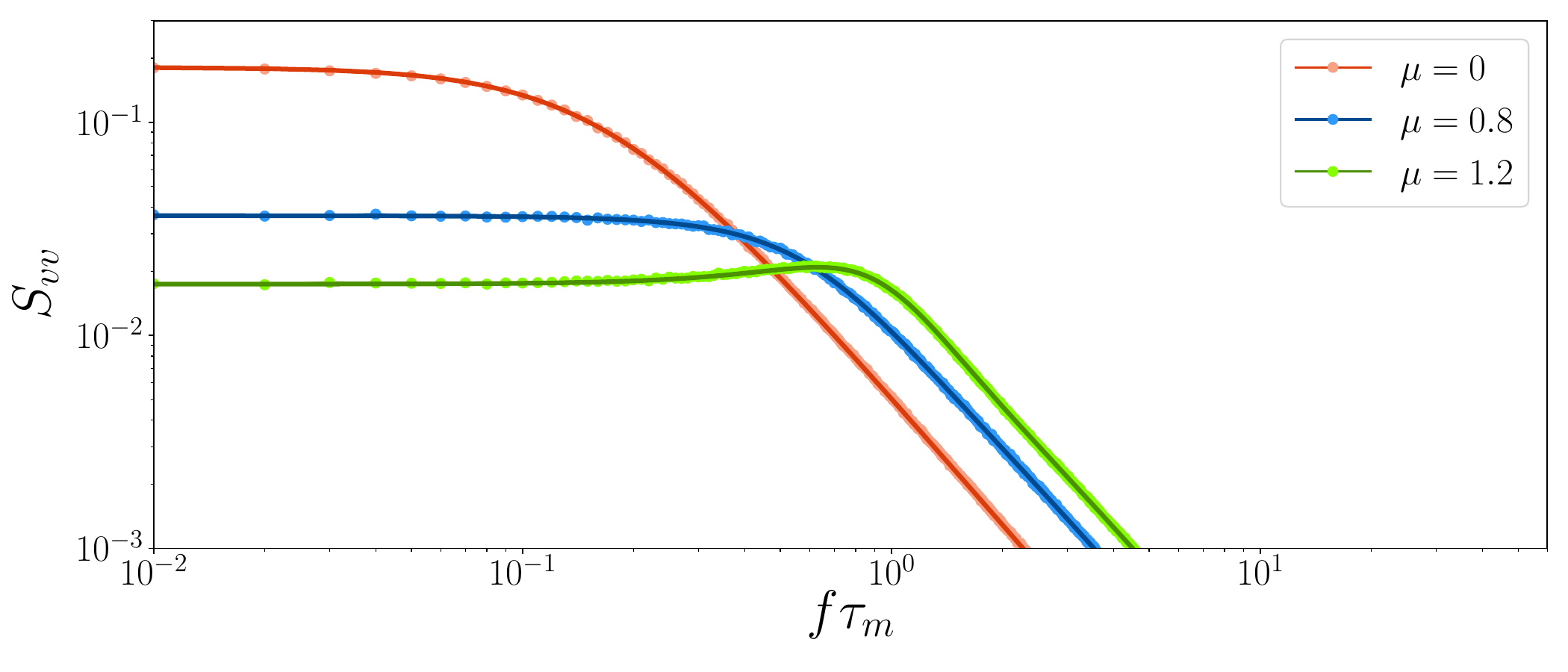}
	\caption{{\bf Numerical validation of Eq.~\eqref{eq:S_vv_theo}} 
    Comparing the prediction of $S_{vv}$ based on Eq.~\eqref{eq:S_vv_theo} (solid lines) and $S_{vv}$ from simulation data (dots) for varying frequency in units of the inverse membrane constant and different mean inputs $\mu$. The noise intensity is $D= 0.1$.}
    \label{fig:S_vv}
\end{figure}

\section{General integrate-and-fire model with adaptation and colored noise}\label{sec:adapt}
We now consider a more general non-refractory model that includes additional important features of neural dynamics, namely, nonlinear subthreshold dynamics, temporally correlated noise, and spike-frequency adaptation:
\begin{align}
    \begin{aligned}
        &\dot{v} = f(v) - a + \eta(t) - (v_T - v_R) x(t)\\
        \tau_a &\dot a = -a + \tau_a \Delta_a x(t).
    \end{aligned}
    \label{eq:Langevin_GIF}
\end{align}
Here, instead of the linear term $\mu - v$ we have a general nonlinear function $f$ and replaced the white Gaussian noise with a temporally correlated (colored) Gaussian noise $\eta(t)$ with correlation function $C_{\eta\eta}(\tau)$ and power spectrum $S_{\eta\eta}(\om)$. Furthermore, the adaptation variable $a(t)$ represents an inhibitory current that drives the membrane voltage away from the threshold. The dynamics of $a$ are given in the second line. In the absence of spikes, $a$ decays exponentially with time constant $\tau_a$. Since adaptation is a slow process, $\tau_a/\tau_m \gg 1$. In this work, we use $\tau_a = 100 \tau_m$ in simulations. A spike causes an instantaneous jump in $a$ of size $\Delta_a$. For a deterministic version of this model, see \cite{BreGer05, TouBre08}.

In this more general model, spike-train susceptibility and spontaneous statistics are also related by an FRR \cite{Lin22}
\be
    \chi_x(\om) = \frac{\left(v_T-v_R+\frac{\tau_a\Delta_a}{1 + i\om\tau_a}\right)S_{xx}(\om) + i\om S_{xv}(\om)-S_{xf}(\om)}{S_{\eta\eta}(\om)}\label{eq:xFRR_adapt},
\ee
where $S_{xf}$ is the cross-spectrum of the spike train and $f$. Adaptation enters the FRR as a frequency-dependent coefficient of $S_{xx}$. The adaptation parameters ($\tau_a$ and $\Delta_a$) cannot be measured directly but accessible through an extra set of step-current experiments \cite{KieLin25}.

\subsection{Fluctuation-response relation for the membrane voltage} 
To derive a FRR for the general IF model with adaptation and colored noise, we can still rely on the Rice-Furutsu-Novikov scheme \cite{Lin22}. Again, we start with \req{Langevin_GIF} in Fourier space, and take the complex conjugate
\begin{align}
        &i\om \Tilde{v}^{\ast}(\om) = \Tilde{f}^{\ast}(\om) - \Tilde{a}^{\ast}(\om) + \Tilde{\eta}^{\ast}(\om) - (v_T - v_R) \Tilde{x}^{\ast}(\om) \notag\\
    &(1+i\tau_a\om)\Tilde{a}^{\ast}(\om) = \tau_a \Delta_a \Tilde{x}^{\ast}(\om).
\end{align}
We can eliminate $\Tilde{a}^{\ast}$ in the first line using the second line such that only one equation remains
\begin{align}
        i\om \Tilde{v}^{\ast}(\om) =& -  \left(v_T - v_R + \frac{\tau_a \Delta_a}{1+i\tau_a\om}\right)\Tilde{x}^{\ast}(\om)\notag\\&+\Tilde{f}^{\ast}(\om) + \Tilde{\eta}^{\ast}(\om).
\end{align}
Next, we multiply by $\Tilde{v}$ and take the average over a stationary ensemble
\begin{align}
            i\om \mean{\Tilde{v}(\om)\Tilde{v}^{\ast}(\om)} =& -  \left(v_T - v_R + \frac{\tau_a \Delta_a}{1+i\tau_a\om}\right)\mean{\Tilde{v}(\om)\Tilde{x}^{\ast}(\om)}\notag\\
            &+\mean{\Tilde{v}(\om)\Tilde{f}^{\ast}(\om)} + \mean{\Tilde{v}(\om)\Tilde{\eta}^{\ast}(\om)}.\label{eq:Langevin_GIF_FT}
\end{align}
As in the case of white noise, the Furutsu-Novikov theorem provides an expression for the average product of voltage and noise that introduces the susceptibility of the subthreshold voltage
\be
    \mean{\Tilde{v}(\om)\Tilde{\eta}^{\ast}(\om)} = \chi_v(\om) \mean{\Tilde{\eta}(\om)\Tilde{\eta}^{\ast}(\om)}.\label{eq:FNT_color}
\ee
Inserting this relation into \req{Langevin_GIF_FT}, and replacing the averages by the corresponding cross- and power spectra yields
\begin{align}
        i\om S_{vv}(\om) =&\, S_{vf}(\om) + \chi_v(\om)S_{\eta\eta}(\om) \notag\\
        &- \left(v_T - v_R + \frac{\tau_a \Delta_a}{1+i\tau_a\om}\right) S_{vx}(\om).
\end{align}
Solving for $\chi_v$ yields the FRR:
\begin{align}
    \chi_v(\om) = \frac{\left(v_T-v_R+\frac{\tau_a\Delta_a}{1 + i\om\tau_a}\right)S_{vx}(\om) + i\om S_{vv}(\om)-S_{vf}(\om)}{S_{\eta\eta}(\om)}\label{eq:vFRR_adapt}.
\end{align}
Apparently, this is similar to \req{xFRR_adapt}: the first index $x$ has simply been replaced by the index $v$ in all cross- and power spectra.
In complete analogy, we can also multiply with $\Tilde{f}$ instead of $\Tilde{v}$ and carry out all the subsequent steps outlined above, yielding 
\begin{align}
    \chi_f(\om) = \frac{\left(v_T-v_R+\frac{\tau_a\Delta_a}{1 + i\om\tau_a}\right)S_{fx}(\om) + i\om S_{fv}(\om)-S_{ff}(\om)}{S_{\eta\eta}(\om)}\label{eq:fFRR_adapt}.
\end{align}
In the following, we content ourself with a numerical test of \req{vFRR_adapt}, using an exponential nonlinearity 
\be
    f(v) = \mu - v + \Delta_v\ \eops^{\frac{v-v_t}{\Delta_v}},
\ee
where $\Delta_v = 0.2$ and $v_t = 1$ characterize the width and transition from linear to exponential behavior. Furthermore, for the colored noise we select a lowpass-filtered Gaussian noise, an Ornstein-Uhlenbeck process with exponential correlation function and Lorentzian power spectrum
\be
    C_{\eta\eta}(\tau) = \sigma^2 \eops^{-\frac{|\tau|}{\tau_c}}, \quad S_{\eta\eta}(\om) = \frac{2 \sigma^2 \tau_c}{1+ (\om \tau_c)^2}.
\ee
Here, $\sigma^2$ and $\tau_c$ are the variance and correlation time of the Ornstein-Uhlenbeck process, respectively.  

In \bi{vFRR_adapt} we demonstrate the validity of \req{vFRR_adapt} for two distinct parameters sets. In panel (\textbf{A}) we use a strong and slow adaptation with $\Delta_a = 1$, resulting in a highpass shape of the susceptibility with reduced transmission of low-frequency stimuli. The estimation of the susceptibility via the spontaneous activity using the FRR provides a smooth curve that agrees well with the susceptibility obtained from a broadband stimulation. Because the firing rate is strongly reduced by the adaptation in this case, the Ornstein-Uhlenbeck approximation \req{susci_OUP} for the susceptibility describes the voltage susceptibility at high frequencies well; the main effect of the adaptation (that is not included in this approximation) dominates at low frequencies. 

In \bi{vFRR_adapt}(\textbf{B}), we use a weak adaptation with $\Delta_a = 0.005$ that still reduces the firing rate substantially (from $r_0(\Delta_a = 0) = 0.26$ to $r_0(\Delta_a = 0.005) = 0.22$). The reduction of transmission at low frequencies is not present anymore and the Ornstein-Uhlenbeck approximation does not fit the susceptibility except at very high frequencies. Nevertheless, the agreement between the directly measured susceptibility and the one estimated from the spontaneous activity is still good. The divergence and deviations at high frequencies are caused by the power spectrum of the noise in the denominator as the power of the Ornstein-Uhlenbeck noise is strongly reduced at high frequencies for the chosen parameters. 

\begin{figure}[ht] 
	\includegraphics[width= 0.484\textwidth,angle=0]{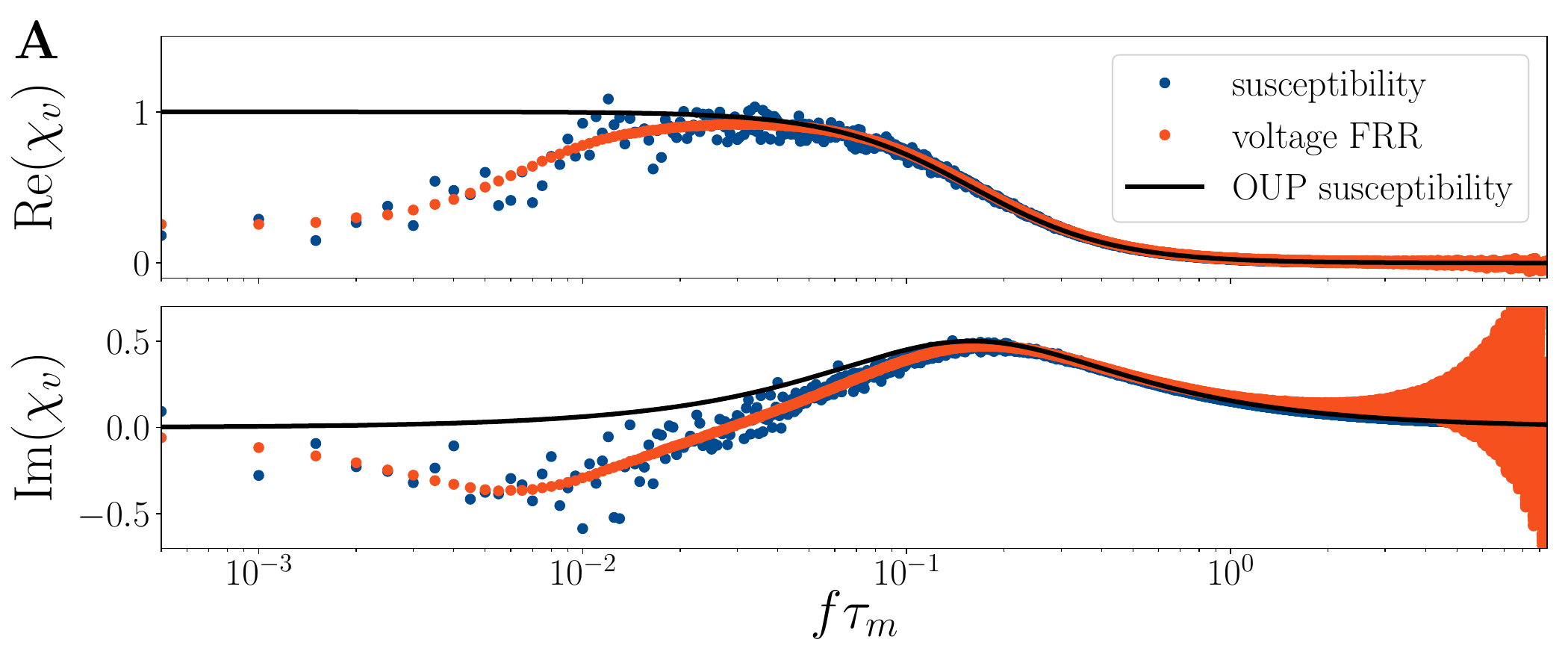}
    \includegraphics[width= 0.484\textwidth,angle=0]{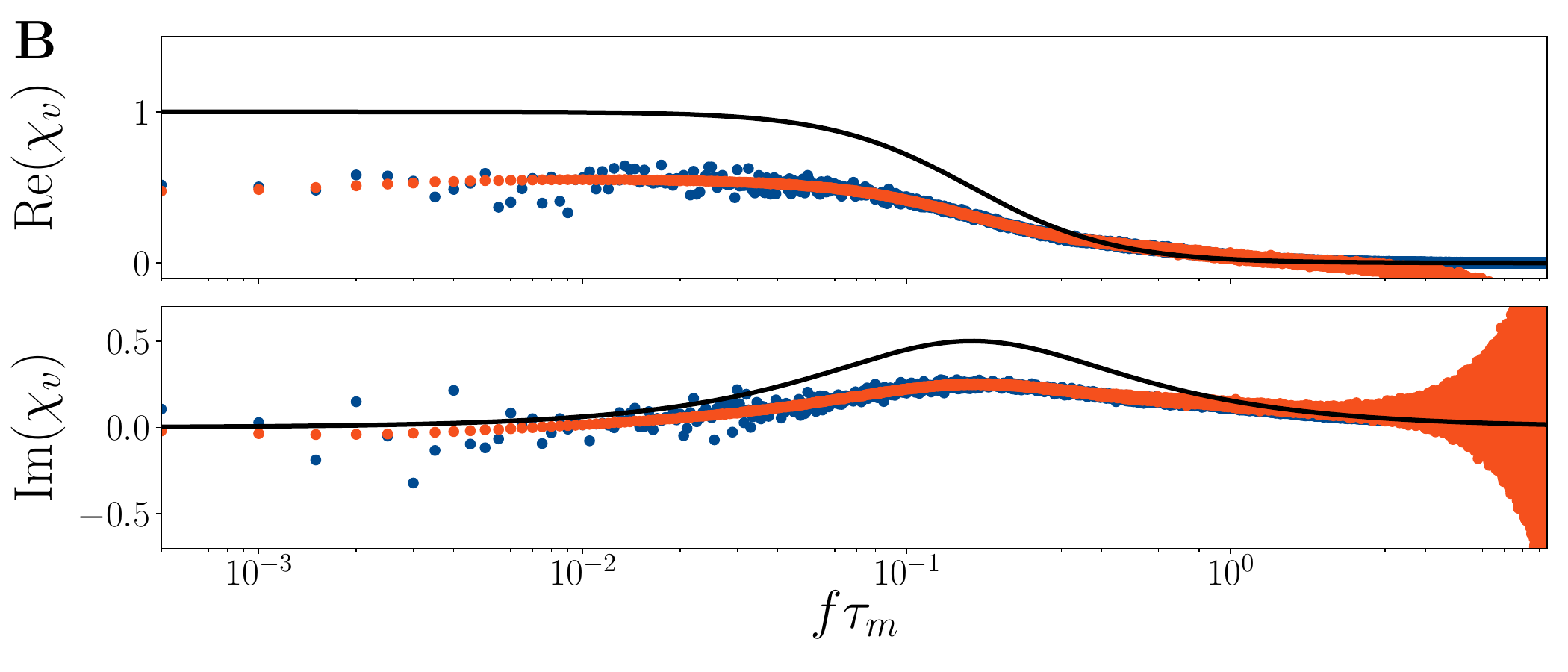}
	\caption{{\bf Numerical validation of the voltage FRR \req{vFRR_adapt}.} 
    Real and imaginary part separately shown for strong adaptation $\Delta_a = 1$ (\textbf{A}) and weak adaptation $\Delta_a = 0.005$ (\textbf{b}). Comparing response statistics and fluctuation statistics, i.e. left and right part of \req{vFRR_adapt}, for varying frequency in units of the inverse membrane constant with $\mu = 0.8$, $v_R = 0$, $v_T = 1.5$, time step of $\Delta t = 5\cdot 10^{-5}$, $N = 4\cdot 10^{4}$ realizations and a total simulated time of $T = 2000$. Colored noise is realized by an Ornstein-Uhlenbeck process with correlation time $\tau_{\text{OUP}} = 10$ and variance $\sigma^2 = 0.5$. The response was determined from stimulation with a broadband noise $\varepsilon s(t)$ with power in the range $[-100, 100]$, scaled such that $s(t)$ has unit variance and $\varepsilon = 1$. Solid black line shows the response of the Ornstein-Uhlenbeck process \req{susci_OUP}.}
    \label{fig:vFRR_adapt}
\end{figure}

\subsection{Response-response relation for membrane voltage, voltage nonlinearity, and spike train}
For the general IF model with adaptation and Gaussian colored noise, we can derive response-response relations (RRR) between different susceptibilities too. Specifically, for a stimulus in the form of a complex exponential, the linear response in the various mean values define the susceptibilities $\chi(\om)$ as follows  
\ba
    \begin{aligned}
        \mean{v(t)} &= \lr{v}_0 + \epi\eops^{-i\om t}\chi_v(\om)\\
        \mean{x(t)} &= r_0 + \epi\eops^{-i\om t}\chi_x(\om)\\
        \mean{f(v)} &= \lr{f}_0 + \epi\eops^{-i\om t}\chi_f(\om)\\
        \mean{a(t)} &= \lr{a}_0 + \epi\eops^{-i\om t}\chi_a(\om).
    \end{aligned}
    \label{eq:suscis}
\ea
In order to calculate the RRR, we apply the generalization of method (ii) of sec.~\ref{sec:RRR} and perform an ensemble average of \req{Langevin_GIF}
\begin{align}
    &\frac{\d}{\d t} \mean{ v(t) }  = \mean{ f(v) } -\mean{ a(t) } - (v_T - v_R) \mean{ x(t)} + \epi \eops^{-i\om t}\notag\\
     \tau_a &\frac{\d}{\d t}\mean{ a(t)} = -\mean{ a(t) } + \tau_a \Delta_a \mean{ x(t)}\label{eq:Langevin_adapt_RRR},
\end{align}
where the zero-mean noise drops out. Upon insertion of \req{suscis} into these equations, we obtain two equations for the stationary mean values (zeroth order in $\varepsilon$)
\be
    \lr{f}_0=\lr{a}_0 + (v_T - v_R)r_0,\quad    \lr{a}_0 =\, \tau_a \Delta_a r_0,
\ee
relations that can be easily used to express $\lr{f}_0$ by the firing rate (without using the inaccessible mean value of the adaptation variable). The first order in $\varepsilon$ yields two more equations, that can be combined to obtain a relation between the susceptibilities 
\begin{align}
    \chi_x(\om) = \frac{1+ i\om\chi_v(\om) +\chi_f(\om)}{v_T-v_R +\frac{\tau_a\Delta_a}{1-i\om\tau_a}}\label{eq:RRR_adapt}.
\end{align}
\begin{figure}[ht] 
	\includegraphics[width= 0.484\textwidth,angle=0]{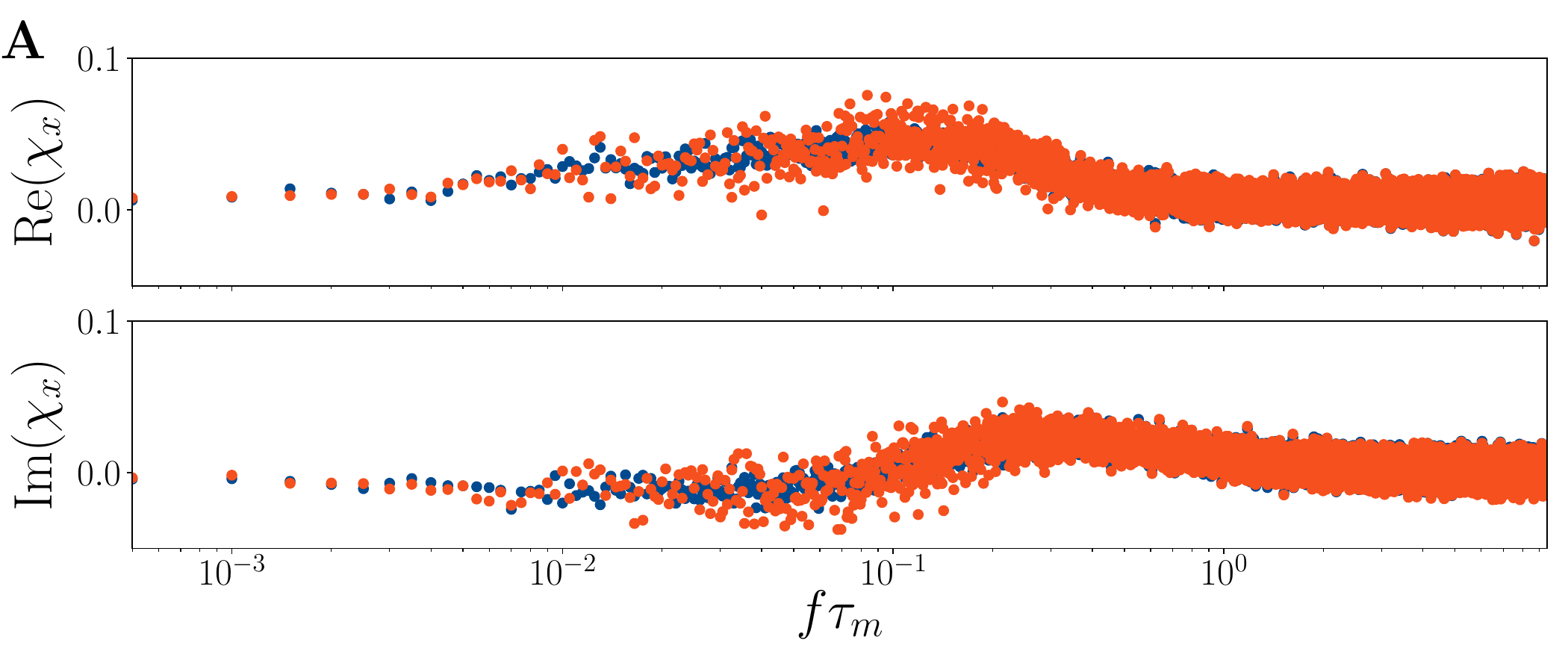}
    \includegraphics[width= 0.484\textwidth,angle=0]{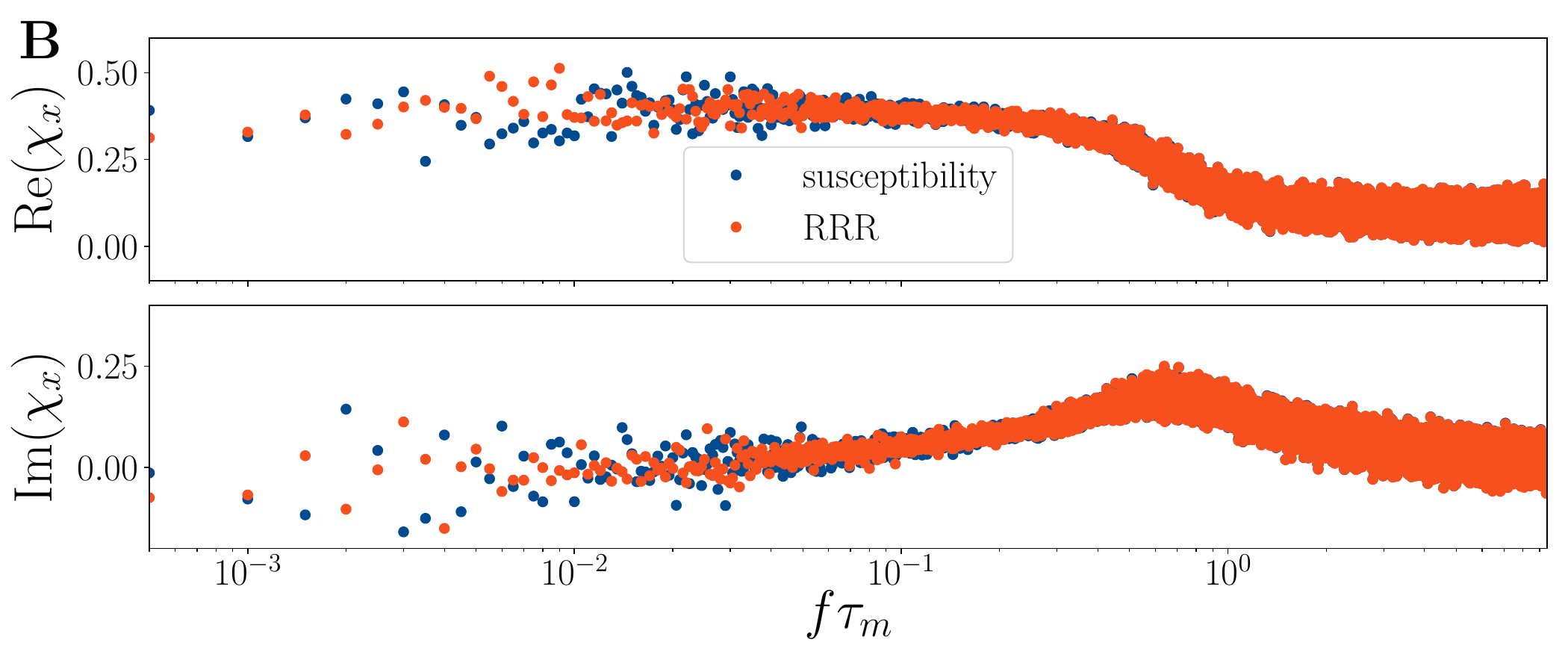}
	\caption{{\bf Numerical validation of the RRR \req{RRR_adapt}.} 
    Real and imaginary part separately shown for strong adaptation $\Delta_a = 1$ (\textbf{A}) and weak adaptation $\Delta_a = 0.005$ (\textbf{b}). Comparing direct measurement of $\chi_x$ and estimate from \req{RRR_adapt} for varying frequency in units of the inverse membrane constant with $\mu = 0.8$, $\tau_a = 100$, $v_t = 1$, $\Delta_v = 0.2$, $v_R = 0$, $v_T = 1.5$, time step of $\Delta t = 5\cdot 10^{-5}$, $N = 4\cdot 10^{4}$ realizations and a total simulated time of $T = 2000$. Colored noise is realized by an Ornstein-Uhlenbeck process with correlation time $\tau_{\text{OUP}} = 10$ and variance $\sigma^2 = 0.5$. The response was determined from stimulation with a broadband noise $\varepsilon s(t)$ with power in the range $[-100, 100]$, scaled such that $s(t)$ has unit variance and $\varepsilon = 1$.}
    \label{fig:RRR_adapt}
\end{figure} 
This relation is verified in \bi{RRR_adapt} for the two parameter sets that we used already in \bi{vFRR_adapt}. Here, we can use the same simulation for estimation of the left hand side and the right hand side; we find a good agreement for both parameter sets. 

\subsection{Power spectrum of the subthreshold membrane voltage}
We can obtain a generalized version of \req{S_vv_theo} accounting for the general function $f$, the adaptation current, and the colored noise. We combine the FRR for the response of the spike train \req{xFRR_adapt}, the FRR for the response of the subthreshold membrane potential \req{vFRR_adapt}, the FRR for the response of $f$ \req{fFRR_adapt}, and the RRR \req{RRR_adapt}, yielding the relation
\begin{align}
    &\om^2S_{vv}(\om) - 2\om\text{Im}S_{vf}(\om) + S_{ff}(\om) = \notag\\ 
    &(v_T - v_R)^2 S_{xx}(\om) + S_{\eta\eta}(\om)[ 1 - 2(v_T -v_R)\text{Re}\chi_x(\om)] \notag\\
    &+ \frac{\tau_a\Delta_a}{1 + \om^2\tau_a^2}\{[2(v_T-v_R) + \tau_a\Delta_a]S_{xx}(\om) \notag\\ 
    &- 2 S_{\eta\eta}(\om)[\text{Re}\chi_x(\om)-\tau_a\om\text{Im}\chi_x(\om)]\}.\label{eq:S_vv_adapt}
\end{align}
When considering specifically the LIF model, $f(v) = \mu - v$, the FRR \req{vFRR_adapt} is given by
\begin{align}
    \chi_v(\om) = \frac{\left(v_T-v_R+\frac{\tau_a\Delta_a}{1 + i\om\tau_a}\right)S_{vx}(\om) + (1 + i\om)S_{vv}(\om)}{S_{\eta\eta}(\om)}\label{eq:vFRR_LIF_adapt},
\end{align}
and the RRR \req{RRR_adapt} simplifies to 
\begin{align}
    \chi_x(\om) = \frac{1- (1-i\om)\chi_v(\om)}{v_T-v_R +\frac{\tau_a\Delta_a}{1-i\om\tau_a}}\label{eq:RRR_LIF_adapt}.
\end{align}
Further, we obtain a relation for the power spectrum of the subthreshold membrane potential
\begin{align}
    &S_{vv}(\om) = \notag\\ 
    &\frac{(v_T - v_R)^2 S_{xx}(\om) + S_{\eta\eta}(\om)[ 1 - 2(v_T -v_R)\text{Re}\chi_x(\om)]}{1 + \om^2} \notag\\
    &+ \frac{\tau_a\Delta_a}{(1 + \om^2\tau_a^2)(1+\om^2)}\{[2(v_T-v_R) + \tau_a\Delta_a]S_{xx}(\om) \notag\\ 
    &- 2 S_{\eta\eta}(\om)[\text{Re}\chi_x(\om)-\tau_a\om\text{Im}\chi_x(\om)]\}.\label{eq:S_vv_LIF_adapt}
\end{align}
This expresses the power spectrum solely by the spontaneous and response statistics of the spike train and the noise spectrum. The second line resembles the original relation \req{S_vv_theo}; only the noise intensity is replaced by $S_{\eta\eta}$. The third and fourth line correspond to the contributions to the power spectrum due to adaptation. We tested and confirmed \req{S_vv_adapt} and \req{S_vv_LIF_adapt} (results not shown).

\section{Summary and Conclusions}\label{sec:diss}
In this work, we have established various relations linking spontaneous and response statistics of subthreshold membrane potential and spike train of different IF models, see Tab.~\ref{tab:overview}. Following the methods from \cite{Lin22}, we have found fluctuation-response relations (FRR) for the response of the subthreshold voltage, response-response relations (RRR) between voltage and spike-train responses, and expressions for the power spectrum of the voltage independent of other voltage statistics.
\begin{table*}[ht]
    \begin{tabular}{ |c|c|c|c| } 
        \hline
        model & voltage FRR & RRR & voltage power spectrum \\ 
        \hline
        LIF model with Gaussian white noise & \req{vFRR} & \req{RRR}$^{\ast, \dagger}$ & \req{S_vv_theo}$^{\ast}$ \\ 
        \hline
        LIF model with Gaussian white noise and refractory period & / & \req{RRR_ref}$^{\ast}$ & /\\ 
        \hline
        general IF model with adaptation and Gaussian colored noise & \req{vFRR_adapt} & \req{RRR_adapt}$^{\dagger}$ & /\\ 
        \hline
        LIF model with adaptation and Gaussian colored noise & \req{vFRR_LIF_adapt} & \req{RRR_LIF_adapt} & \req{S_vv_LIF_adapt}\\ 
        \hline
    \end{tabular}
    \caption{\textbf{Overview of derived relations.} $^{\ast}$ indicates a new analytical result, and $^{\dagger}$ marks results that also hold for non-Gaussian noise.}
     \label{tab:overview}
\end{table*}

The derived relations might be useful for various applications. First, the FRRs could be used like in statistical physics: estimating the response characteristics based on the spontaneous fluctuations. Second, assuming a certain neuron is described by a specific IF model, the derived relations propose new criteria by which to judge the validity of this choice of model. Thirdly, similar to the FRR for the response of the spike train \cite{Lin22}, one could estimate the power spectrum of the intrinsic colored noise using the FRR for the voltage and the relation for the power spectrum of the voltage. Fourthly, the RRR could be used to estimate the response statistics of the membrane voltage from the spike-train susceptibility alone; notably, the RRR does not require knowledge of the noise statistics. Finally, the relation for the power spectrum of the voltage can be likewise leveraged to estimate the voltage spectrum from spike-train statistics; here, the knowledge of the intrinsic noise level is required, which can be, at least for the LIF model with white Gaussian noise, uniquely determined from firing rate and coefficient of variation of the spike train \cite{VilLin09}.

An open problem is the extension of the RRR in the refractory case to account for the voltage behavior during the spike by means of the Rice-Furutsu-Novikov method as employed in \cite{PutLin24} and to obtain an FRR for the response of the voltage in the same way. Recently, a Furutsu-Novikov-like relation for shot noise has been found and used to derive an FRR for the response to a rate-modulated stimulus \cite{StuLin24}. In combination with the RRRs derived in this work, that also hold for non-Gaussian shot noise, the presented framework could be extended. The Rice-Furutsu-Novikov method used in our derivations could also be applied to models with additional variables besides voltage and adaptation, for instance to multi-compartment models or models with a bursting mechanism. 
 
\appendix
\section{Details of the simplification of \req{I_def}}\label{sec:Appen}
To calculate $I(v_\ell)$, given in \req{I_def}, we break down the full expression into smaller pieces that we consider separately
\begin{align}
   I(v_\ell) =& \frac{1}{Dq_1(v_T)} \bigg[q_1(v_\ell)q_2(v_T)J_1(v_T)\notag\\ 
   &-q_1(v_T)q_2(v_\ell) J_1(v_\ell) - q_1(v_\ell)q_1(v_T)J_2(v_\ell)\bigg]\label{eq:I},
\end{align}
with
\begin{align}
    \begin{aligned}
        &J_1(v_k):=\int_{-\infty}^{v_k} \d v\, v\, \eops_-(v) q_1(v) \quad,\ k = T, \ell\\
        &J_2(v_\ell):=\int_{v_\ell}^{v_T} \d v\, v\, \eops_-(v) q_2(v).\label{eq:J_1_J_2}
    \end{aligned}
\end{align}
The integrands of these integrals differ only in the occurring parabolic cylinder functions $q_1$, $q_2$ (see \req{U_V}) leading to similar steps in the calculation. Thus, we only present the calculation of $J_1$ in detail and comment on deviations in the calculation of $J_2$. 

The first step is an integration by parts:
\begin{align}
    J_1(v_k) =&  v\, Q_1(-1/2-i\om,v) \Big|_{-\infty}^{v_k} - \int_{-\infty}^{v_k} \d v\, Q_1(-1/2-i\om,v)\label{eq:PCF_IbyP},
\end{align}
where $Q_1(-1/2-i\om,v)$ is the antiderivative of $\eops_-(v) U\left(-1/2-i\om, z_v \right)$ (here, $z_u = (\mu - u)/\sqrt{D}$), i.e. $d Q_1(-1/2-i\om,v) / dv = \eops_-(v) U\left(-1/2-i\om, z_v \right)$. Using the recurrence relation (see (19.6.2) in \cite{AbrSte70})
\be
    \frac{\d}{\d z}U(a,z) - \frac{z}{2}U(a,z) + U(a-1,z) = 0,
\ee
and the product rule for differentiation of $\eops_-(v) U\left(a, z_v \right)$ with respect to $v$, it is straightforward to show that
\be
    Q_1(a,v) =  \sqrt{D}\, U\left(a+1, z_v\right) \eops_-(v).\label{eq:PCF_int_U}
\ee
Analogously for the antiderivative $Q_2(a,v)$ of $\sqrt{\pi/2}\, \eops_-(v)V\left(-1/2-i\om, z_v \right)$ in $J_2$, by using the recurrence relation (see (19.6.5) in \cite{AbrSte70})
\be
    \frac{\d}{\d z}V(a,z) - \frac{z}{2}V(a,z) - \left(a-\frac{1}{2}\right) V(a-1,z) = 0,
\ee
we find 
\be
    Q_2(a,v) =  -\frac{\sqrt{D}}{a+\frac{1}{2}}\, V\left(a+1, z_v\right) \eops_-(v).\label{eq:PCF_int_V}
\ee

Applying \req{PCF_int_U} twice to \req{PCF_IbyP}, we obtain
\begin{align}
    J_1(v_k) =&\sqrt{D}\Big[v_j\, \eops_-(v_k) q_{11}(v_k) - \sqrt{D}\, \eops_-(v_k) q_{12}(v_k) \notag\\
    &- \lim_{v \to - \infty}v\, \left\{\eops_-(v) q_{11}(v)  + \sqrt{D}\, \eops_-(v) q_{12}(v)\right\}\Big]  \label{eq:J_1},
\end{align}
where $q_{11}$ and $q_{12}$ are parabolic cylinder functions with an increased first argument
\begin{align}
    \begin{aligned}
        q_{11}(v) &= U\left(\frac{1}{2} - i\om, \frac{\mu - v}{\sqrt{D}}\right), \\
        q_{12}(v) &= U\left(\frac{3}{2} - i\om, \frac{\mu - v}{\sqrt{D}}\right).
    \end{aligned}
\end{align}
Using an expansion of $U(a,z)$ for $z>0$ and $z \gg |a|$ (see (19.8.1) in \cite{AbrSte70})
\be
    U(a,z) \sim \mathrm{exp}\left[ -\frac{1}{4}z^2\right]z^{-a-\frac{1}{2}}\left[1-\mathcal{O}(z^{-2})\right],
\ee
we see that the limit $v \to -\infty$ (second line in \req{J_1}) vanishes. Following the same steps for the second line of \req{J_1_J_2} with \req{PCF_int_V} instead of \req{PCF_int_U} yields
\begin{align}
    J_2(v_\ell) =& \frac{\sqrt{D}}{i\om}\bigg[v_T\, \eops_-(v_T) q_{21}(v_T) +\frac{\sqrt{D}}{1-i\om}\eops_-(v_T) q_{22}(v_T)\notag\\
    &- v_\ell\, \eops_-(v_\ell) q_{21}(v_\ell)  - \frac{\sqrt{D}}{1-i\om}\eops_-(v_\ell) q_{22}(v_\ell) \bigg],\label{eq:J_2}
\end{align}
with parabolic cylinder functions
\begin{align}
    \begin{aligned}
        q_{21}(v) &=\sqrt{\frac{\pi}{2}}\, V\left(\frac{1}{2} - i\om, \frac{\mu - v}{\sqrt{D}}\right), \\
        q_{22}(v) &=\sqrt{\frac{\pi}{2}}\, V\left(\frac{3}{2} - i\om, \frac{\mu - v}{\sqrt{D}}\right).
    \end{aligned}
\end{align}
Inserting \req{J_1} and \req{J_2} back into \req{I}, we obtain
\begin{align}
          I(v_\ell)=& \frac{\eops_-(v_T)}{\sqrt{D}}\frac{q_1(v_\ell)}{q_1(v_T)}\left[v_T A_1(v_T) - \sqrt{D}\ A_2(v_T) \right]\notag\\ &- \frac{\eops_-(v_\ell)}{\sqrt{D}}\left[v_\ell A_1(v_\ell)-\sqrt{D}\ A_2(v_\ell)\right],\label{eq:I_A_1_A_2}
\end{align}
where
\begin{align}
    A_1(v_k) =& q_2(v_k)q_{11}(v_k) -\frac{q_{21}(v_k)q_1(v_k)}{i\om}\\
    A_2(v_k) =& q_2(v_k)q_{12}(v_k) + \frac{q_{22}(v_k)q_1(v_k)}{i\om(1-i\om)}.
\end{align}
Next, we want to simplify $A_1$ and $A_2$. To achieve this, we apply some additional recurrence relations (Eqs.~(19.6.1),~(19.6.6) in \cite{AbrSte70})
\begin{align}
    \begin{aligned}
        \frac{\d}{\d z}U(a,z) &+ \frac{z}{2}U(a,z) + \left(a+\frac{1}{2}\right)U(a+1,z) = 0, \\
        \frac{\d}{\d z}V(a,z) &+ \frac{z}{2}V(a,z) - V(a-1,z) = 0,
    \end{aligned}  
    \label{eq:Recu_UV}
\end{align}
as well as another property of the parabolic cylinder functions 
\be
    U(a,z)\frac{\d}{\d z_v}V(a,z_v) - V(a,z_v)\frac{\d}{\d z_v}U(a,z_v) = -\sqrt{\frac{2D}{\pi}}\label{eq:UV_prod}.
\ee
This follows from the Wronskian of these parabolic cylinder functions (Eq.~(19.4.1) in \cite{AbrSte70}) and the application of the chain rule. We use \req{UV_prod} to eliminate the parabolic cylinder functions from $A_1$ and $A_2$. To do so, we first have to bring all parabolic cylinder functions to the same first argument. We achieve this by applying the recurrence relations \req{Recu_UV} for $A_1$ once and for $A_2$ twice. Doing all this yields
\begin{align}
    \begin{aligned}
        A_1 =& \frac{\sqrt{D}}{i\om}\\
        A_2(v_k) =&\frac{v_k-\mu}{i\om(1-i\om)}.\label{eq:A_1_A_2}
    \end{aligned}
\end{align}
By inserting \req{A_1_A_2} into \req{I_A_1_A_2}, we end up with the final result for $I$ 
\begin{align}
    I(v_\ell) =& \frac{1}{i\om(1-i\om)}\bigg[\eops_-(v_T)\frac{q_1(v_\ell)}{q_1(v_T)}(\mu-i\om v_T) \notag\\
    &- \eops_-(v_\ell) (\mu-i\om v_\ell)\bigg].
\end{align}

\end{document}